\RequirePackage{fix-cm}
\documentclass[smallextended]{svjour3}       % onecolumn (second format)
\smartqed  % flush right qed marks, e.g. at end of proof
\usepackage[dvipdfmx]{graphicx}
\usepackage{color}
\usepackage[hyphens]{url}
\usepackage{multirow}
\usepackage{listliketab}
\usepackage{amssymb}
\usepackage{fancybox}
\usepackage[table]{xcolor}
\usepackage{colortbl}
\usepackage{caption}
\begin{document}

%\newboolean{showcomments}
%\setboolean{showcomments}{true} % toggle to show or hide comments
%\ifthenelse{\boolean{showcomments}}
%{\newcommand{\nbnote}[2]{
%		% \fbox{\bfseries\sffamily\scriptsize#1}
%		\fcolorbox{blue}{yellow}{\bfseries\sffamily\scriptsize#1}
%		{\sf\small\textit{#2}}
%		% \marginpar{\fbox{\bfseries\sffamily#1}}
%	}
%}
%{\newcommand{\nbnote}[2]{}
%	\newcommand{\version}{}
%}
\newcommand\ikeda[1]{\nbnote{Shohei}{\textcolor{magenta}{#1}}}
\newcommand\akinori[1]{\nbnote{Akinori}{\textcolor{blue}{#1}}}
\newcommand\raula[1]{\nbnote{Raula}{\textcolor{red}{#1}}}

%%% Our own commands
\newcommand{\figref}[1]{Figure~\ref{#1}}
\newcommand{\tabref}[1]{Table~\ref{#1}}
\newcommand{\bibref}[1]{[\ref{#1}]}

\newcommand{\todo}[1]{{\color{red}#1}}
\newcommand{\reply}[1]{{\color{cyan}#1}}
\newcommand{\conclusionbox}[1]{%
	\vspace{1mm}
	\noindent
	\framebox[1.00\textwidth][c]{%
		\parbox[b]{0.95\textwidth}{%
			#1
		}
	}
}
\newcommand{\rmd}{\texttt{README}~}
\newcommand{\package}[1]{\texttt{#1}}
\newcommand{\algorithm}[1]{\texttt{#1}}
\newcommand{\content}[1]{``#1''}
\newcommand{\keyword}[1]{``#1''}

\newcommand{\rqone}{What do developers write in a \rmd file?}
\newcommand{\rqtwo}{Does the type of project affect how developers write their \rmd file?}
\newcommand{\rqthree}{How does the \rmd evolve over time?}    

\newcommand{\bounds}[1]{$#1\%\pm5\%$}
\newcommand{\ea}{\emph{et al.}}
\newcommand{\smallsection}[1]{\noindent\textbf{#1.}}
\newcommand{\dl}{{{Discussion length }}}

\newcommand{\ite}{{{\# iterations }}}
%%%

\title{An Empirical Study on \rmd contents for JavaScript Packages
%How do developers receive timely code reviews based on technical and non-technical factors?
}
\subtitle{}


\author{Shohei Ikeda         \and
        Akinori Ihara	\and%etc.
	Raula Gaikovina Kula  	\and
        Kenichi Matsumoto
}


\institute{Shohei Ikeda \and Akinori Ihara \and Raula Gaikovina Kula \and Kenichi Matsumoto \at
              Software Engineering Laboratory, \\
              Nara Institute of Science and Technology, Japan \\
              \email{ikeda.shohei.ik5, akinori-i@is.naist.jp, raula-k@is.naist.jp, matumoto@is.naist.jp}           %  \\
}

\date{Received: date / Accepted: date}

\maketitle

\begin{abstract}

Contemporary software projects often utilize a \texttt{README.md} to share crucial information such as installation and usage examples related to their software. 
Furthermore, these files serve as an important source of updated and useful documentation for developers and prospective users of the software.
%This is because developers struggle with keeping documentation up to date.
Nonetheless, both novice and seasoned developers are sometimes unsure of what is required for a good \texttt{README} file.
To understand the contents of \rmd, we investigate the contents of 43,900 JavaScript packages.
Results show that these packages contain common content themes (i.e., `usage', `install' and `license'). 
Furthermore, we find that application-specific packages more frequently included content themes such as `options', while library-based packages more frequently included other specific content themes (i.e., `install' and `license'). 
%The study concludes that \rmd files reveal insights reflecting project changes and assists especially novice developers in the project.
\end{abstract}

\newpage

%%% Section 1
%%%%%%%%%%%%%%%%%%%%%%%%%%%%%
\section{Introduction} \label{sec:introduction}
%%%%%%%%%%%%%%%%%%%%%%%%%%%%%

To encourage prospective users and interested developers to write documentation, it is common practice for Open Source Software (OSS) projects to release software artifacts (i.e., source code, configuration files and documentation) through platforms such as GitHub.
Some projects release a meta-file document called \texttt{README}, which typically includes a summary of the most useful and updated information, such as an install guide and usage examples.
This is especially crucial for tracking changes once newer versions get released. 
In fact, all GitHub hosted projects present the \rmd on their front page~\cite{Coelho_ESEC2017}.

Developers often struggle to write documentation~\cite{IEEEsoftware2003_Lethbridge}.
A large-scale GitHub survey\footnote{Open Source Survey: \url{http://opensourcesurvey.org/2017/}} conducted in June 2017, reported that although software documentation is highly valued, it is frequently overlooked.
Furthermore, most respondents (approximately 93\%) complained that most documentation is either incomplete or outdated. 
In the survey, 60\% of contributors said that they rarely or never contribute to documentation. 
Related studies also confirm that developers struggle to write documentation.
Abebe et al.~\cite{EMSE2016_Abebe} advised developers to note several content themes such as title, system overview, resource requirements, installation, and addressed issues (i.e., new features, bug fixes, and improvements) as caveats in the release note. 
Moreno et al.~\cite{Moreno_TSE2017} reported that developers find it difficult to summarize a release note because it has several content themes, such as fixed bugs, new features, and the improvement of existing features. 
They proposed an approach to automatically generate release notes. 
Similarly, other works~\cite{TOIS2013_Kim}\cite{ICSE2017_Zhou} investigated the relationship between source code (i.e., API, code examples) and documentation.
In terms of \rmd files, Hassan et al.~\cite{ICSE2017_Hassan} proposed an approach to extract a build command, while Zhang et al.~\cite{SANER2017_Zhang} used this approach to identify systems with similar functions. 

A \rmd file contains key documentation patterns for developers, especially when uncovering documentation patterns specific to the types of software.
For instance, {library-specific projects} (i.e., projects used by other applications as third-party libraries) may write their \rmd file differently in {application-specific projects} (i.e., projects used by end-users).

In this study, we would like to understand the extent to which developers write and maintain their \rmd files.
%Our goal is assist novice developers with writing and maintaining a good \rmd file.
We conduct an empirical case study that analyzes over 43,900 packages belonging to the npm JavaScript ecosystem in GitHub.
In particular, we investigate (i) what constitutes typical content themes and (ii) whether content themes indicate the type of a package (i.e., library-specific vs. application-specific).
In this novel study, we learned the following valuable lessons along the way:

\begin{itemize}
\item\textbf{Lesson 1:} \textbf{It is useful to build and summarize a taxonomy of 22 \rmd content themes, which are used by more than 1\% of packages.} -
From over 30,000 content variations, we used a semi-automatic method to build a taxonomy of \rmd content themes. 
%\vspace{2mm}
\item\textbf{Lesson 2: \content{Usage}, \content{Install}, and \content{License} are common \rmd content themes.} - This result complements known guidelines for writing good documentation. 
We also found that less apparent \rmd content themes include \content{API}, \content{Test}, and \content{Todo}, are used in 10\%-24\% of packages.
%\vspace{2mm}
\item\textbf{Lesson 3:} \textbf{Our study shows that \content{Install} and \content{License} are likely content themes for library-specific packages, while the \content{Option} content theme is more common for application-specific packages.} - \content{Install} (i.e., 40\% packages) and \content{License} (i.e., 20\% packages) are common for npm libraries, while nodejs application packages included the option content themes (i.e.,  10\% packages).
\end{itemize}

We conclude that \rmd files reveal insights such as project practices and product changes.
Such information especially assists especially the novice developer.

This paper is laid out as follows. 
Section~\ref{sec:background} describes the background and  motivation of this study. 
Section~\ref{sec:dataset} provides the dataset to conduct our empirical study. 
Section~\ref{sec:rq} presents answers to each of the two research questions proposed in this study. 
Section~\ref{sec:discussion} discusses our findings. Section~\ref{sec:threats_to_validity} presents threats to validity. 
Finally, Section~\ref{sec:conclusion} concludes the paper and presents our future work.

%%% Section 2
%%%%%%%%%%%%%%%%%%%%%%%%%%%%%
\section{Motivation \& Research Overview} \label{sec:background}
%%%%%%%%%%%%%%%%%%%%%%%%%%%%%

\subsection{Illustrative Example \& Key Assumptions} \label{subsec:motivating_example}
Co-founder of GitHub, Tom Preston-Werner recently highlighted the importance of the \rmd file, coining Readme Driven Development (RDD)\footnote{Readme Driven Development: \url{http://tom.preston-werner.com/2010/08/23/readme-driven-development.html}} as an important subset of Document Driven Development.
In this paper, our motivation is to investigate the following assumptions:
\begin{itemize}
%\item \rmd files are a reliable source of up--to--date documentation with important content themes.
\item \rmd file is a reliable source of important documentation changes and content themes of the project.
\item \rmd content themes follow some useful guidelines and may be indicative of its project type.
\end{itemize}

\begin{figure}[t]
\centering
  \includegraphics[width=11cm]{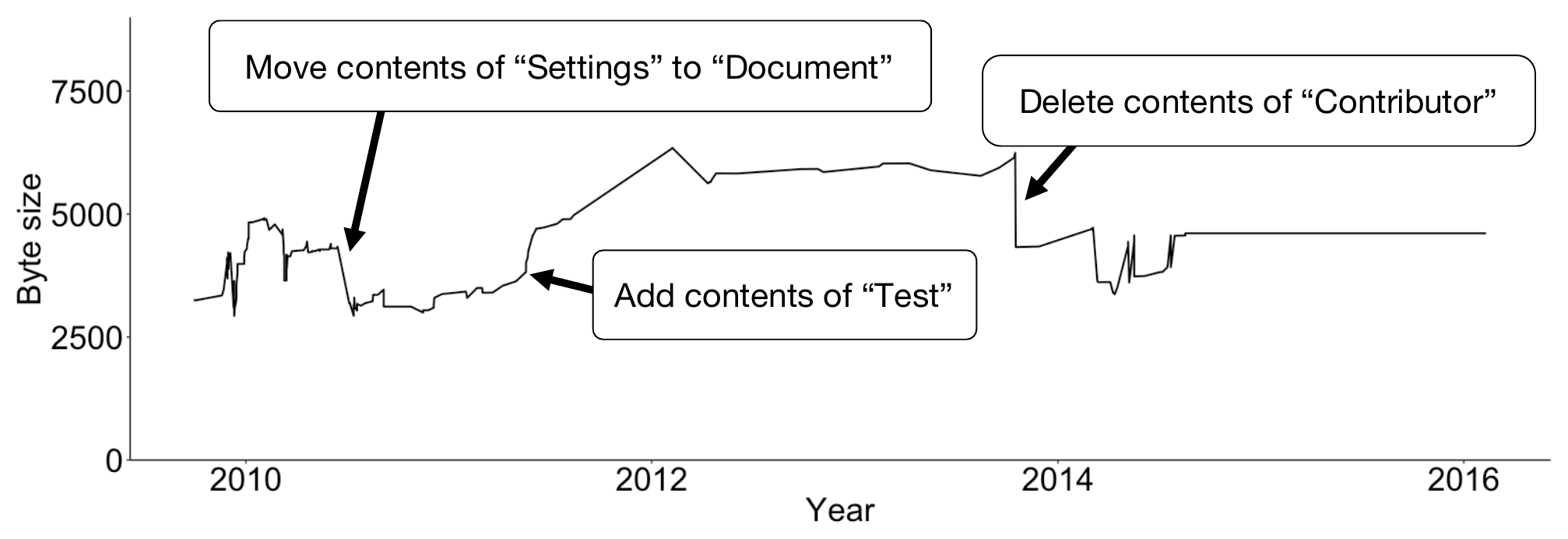}
  \caption{Illustrative example of evolving \rmd file for the \package{express} package. Note that we measure the file size (in bytes) as a measure of changes. } \label{fig:changing_readme}
\end{figure}

Figure~\ref{fig:changing_readme} illustrates an example of how a \rmd changes over time and is indicative of other changes. 
This example shows the JavaScript \package{express}\footnote{express: \url{https://www.npmjs.com/package/express}} package. 
In detail, \package{express} added content themes of \content{Test} in 2011.
For example, \package{express} moved the content theme of \content{Settings} to \content{Documentation} linking to the official website in 2010.
Later, they deleted the content theme of \content{Contributor}.
Interestingly, in a preliminary exploration of 119,093 npm packages, we found on average that a \rmd was updated up to 7 times.

Table~\ref{tab:guideline} shows the existing guidelines that hint at the content theme. 
These guidelines are taken from the following sources: 
\begin{itemize}
\item \textit{GitHub\footnote{GitHub Help -About READMEs-: \url{https://help.github.com/articles/about-readmes/}}} project introduces the content themes which the \rmd file typically includes.
\item \textit{18F\footnote{18F Open Source Style Guide: \url{https://open-source-guide.18f.gov/making-readmes-readable/}}} project is a digital service agency which introduced ``Making \texttt{READMEs} readable''.
\item \textit{OSCON2015\footnote{O'Reilly Open Source Convention: OSCON, July 20 - 24, 2015 in Portland, OR \url{https://conferences.oreilly.com/oscon/open-source-2015}}} is an international conference for open source development. 
Key-note speaker, Mr. Mike Jang explained how open source projects failure to attract users due to poor \rmd quality. 
He later introduced 10 key content themes. 
\end{itemize}

As shown in the Table~\ref{tab:guideline}, we find that key information such as \content{Overview}, \content{Install}, \content{License}, \content{Contribution}, \content{Support}, \content{Author}, \content{Usage}, \content{Release history}, and \content{Product} are perceived as vital for any software projects.

%---------------------------
%---------------------------

%==================================
\begin{table*}[t]
	\centering
	\caption{Guideline for writing a \rmd file}
	\scalebox{0.71}[0.71]{
		\begin{tabular}{l|p{5.5em}|p{5.5em}|p{5.5em}|p{25em}}
			\hline
Content theme&\hfil GitHub \hfil & \hfil 18F \hfil& OSCON2015 \hfil&\hfil Summary \hfil\\
\hline
Overview			&\hfil \checkmark \hfil&\hfil \checkmark \hfil&\hfil \checkmark \hfil	& Description of what the project is for and how useful.\\
Install			&\hfil \checkmark \hfil&\hfil \checkmark \hfil&\hfil \checkmark \hfil	& Instructions for how to develop, use.\\
License			&\hfil \checkmark \hfil&\hfil \checkmark \hfil&\hfil \checkmark \hfil	& List for where your team can ask for contact information.\\
Contribution		&\hfil \checkmark \hfil&\hfil \checkmark \hfil&\hfil \checkmark \hfil	& Instructions for how people can help clarify the documentation.\\
Support			&\hfil \checkmark \hfil&\hfil \checkmark \hfil&\hfil \checkmark \hfil	& List the contact information for your team where to ask questions.\\
Author			&\hfil \checkmark \hfil&&										& List for who maintains and contributes to your project.\\
Usage			&&\hfil \checkmark \hfil&\hfil \checkmark \hfil						& List of code sample and config tips. \\
Release history	&&&\hfil \checkmark \hfil										& List of changes.\\
Product			&\hfil \checkmark \hfil&&										& Description of code for conduct.\\
\hline
		\end{tabular}
	}
	\label{tab:guideline}
\end{table*}
%==================================

%---------------------------
\begin{figure}[p]
\centering
  \includegraphics[width=22cm, angle=90]{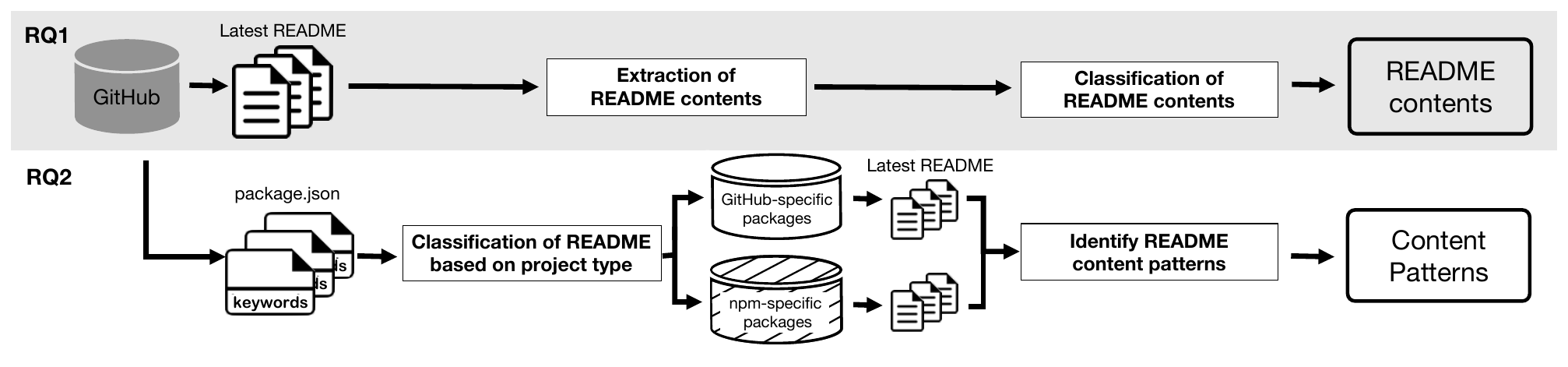}
  \caption{Procedure of this study.} \label{fig:approach}
\end{figure}
%---------------------------

\subsection{Research Questions} \label{subsec:rqs}
Our goal is to understand the content themes.
We therefore investigate the extent to which key content themes (i.e., as found in the guidelines) appear in the \texttt{README}.
As shown in Figure \ref{fig:approach}, we formulate the following research questions to guide our study.

%In this study, we conduct an empirical study of README contents. To achieve this study, we have the following motivations to answer three research questions:
\begin{itemize}
\item \textbf{$RQ_1$: \rqone} As shown in Table~\ref{tab:guideline}, we would like to confirm what content themes constitute a \rmd file. This may provide information for crucial documentation and will confirm the guidelines for good documentation.
%Our goal is to create a taxonomy of \contents.
\item \textbf{$RQ_2$: \rqtwo} We investigate whether or not projects write \rmd files based on their project type.
For example, we speculate that some application frameworks (e.g., \package{express}) or the plugin tools (e.g., \package{gulp}) projects would prioritize purpose and usage of the software (e.g., sample code, example of option). 
%On the other hand, core utility package projects with low-layer functions such as \package{buffer}, testing and benchmark tools. 
\end{itemize}

To answer our research questions, we perform an empirical study on real-world projects. 
Specifically, we perform a case study.
For $RQ_1$, we first extract and conduct a content theme analysis of \rmd files.
Then, for $RQ_2$, we investigate the relationship between the type of project and the content themes.

%
%
%記述すべき項目の指針が求められている．
%そこで本論文では，プロジェクトが最終的に何を記述するのか調査する（RQ1: What is the headline content?）．記述項目の試行錯誤が終了した（安定した）READMEを対象に，記述項目を抽出し，頻繁に記述される項目を明らかにする．
%
%ただし記述項目は，プロジェクトの種類に応じて異なると考えられる．例えばアプリケーションフレームワーク（e.g. express）やアプリケーション構築用のプラグイン支援ツール（e.g. gulp）といったエンドユーザーパッケージでは，設計理念や使用方法（サンプルコードやオプション例）が重視される．一方でバッファの変換（e.g. buffer）や変数の型判別（e.g. kind-of）といった低レイヤーの機能を提供するコアユーティリティパッケージでは，テスト方法やベンチマークが重視される．
%%ただし記述項目は，プロジェクトの種類ごとに異なる．例えばコマンドラインツールではコマンド一覧や通常のインストール方法以外のビルド方法を重視して記述するが，ブラウザで利用するライブラリではサンプルコードや動作可能な環境を重視して記述する．
%記述すべき項目の指針は，プロジェクトの種類ごとに異なるべきである．
%そこでプロジェクトの種類がどの程度記述項目に影響するのか調査する（RQ2-A: How much the keyword impact the headline content?）．
%加えて，記述すべき項目の指針の提案に向けて，RQ2-Aの結果，分類ごとに記述されにくい項目について，その理由を調査する（RQ2-B: Why do not the project use the headline content?）．
%
%また記述項目のうち，頻繁に追加あるいは削除される項目は何か調査し，記述する必要のない項目や，記述することが難しいと考えられる項目を知るための知見を得る．
%\todo{優先度中：RQ3 READMEの項目はどれくらい追加・削除されているか？}

%%% Section 3
%%%%%%%%%%%%%%%%%%%%%%%%%%%%%
\section{Data Collection} \label{sec:dataset}
%%%%%%%%%%%%%%%%%%%%%%%%%%%%%
% I add a sentence just to bridge between 3 and 3.1
In this section, we describe both the data extraction and data preprocessing method for the empirical study.
The final dataset will be used to evaluate the two research questions.
%安定したプロジェクトを使って調べたい．
%なので，そのDatasetはこんな風に用意します．
%\todo{優先度低：文と表で説明}
\subsection{Data extraction}
%\todo{please refer to the npm paper and how they discuss data extraction}
Our study targets a \rmd file for JavaScript packages.
Specifically, we target on JavaScript projects that belong to the npm ecosystem \footnote{npm: ~\url{https://www.npmjs.com/}}, consisting of packages that run on the nodeJS platform.
Packages include libraries (e.g., \package{react}), frameworks (e.g., \package{express}), command line tools (e.g., \package{browserify}), and plug-in-supporting tools to build applications (e.g., \package{grunt}, \package{glup}). 
To support searching these packages, the npm repository adds a keyword tag to each package to explain the features of the package. 
For example, the \package{express} package has \keyword{framework}, \keyword{web} and \keyword{express}.

We use the same process described by Wittern et al.~\cite{Wittern_MSR2016} to extract similar datasets for libraries and applications. 
In detail, we query the npm registry\footnote{accessible on July 1-15, 2016 at \url{https://registry.npmjs.org/-/all}} for all npm packages that were hosted and available on GitHub. 
We then extract the \rmd files for all the packages. 
Finally, we collected 153,857 \rmd files as shown in Table \ref{tab:filtered_packages}.

As shown in Figure~\ref{fig:approach}, our extracted dataset also consists of an extraction of the project type.
Hence, for each project, we extract tagged keywords from the package.json meta-file.
In the Wittern et al. study~\cite{Wittern_MSR2016}, the authors showed that keywords may be indicative of project type, which is needed to answer $RQ_2$.

\subsection{Data Preprocessing}
To ensure the quality of the dataset, we perform filtering to remove the noise in the dataset.
%Our study investigates the \rmd files hosted in GitHub to understand the content themes and the history of the file updates. 
After data extraction, we are able to collect 153,857 \rmd files.
Since most \rmd files are written in the markdown format and we plan to use English as our main language of analysis, we use the \algorithm{unicodedata} library to filter out files which did not use the markdown format (i.e., we exclude files written in Japanese, Cyrillic, or Hangul). Then, there are 141,933 \rmd files. 
Furthermore, to ensure that we capture all initial commits, we only include projects that were created within the three year period of our analysis (i.e., 2013$\sim$2016).
Hence, we left 43,911 \rmd files after preprocessing.

For $RQ_2$, we further separate our dataset into GitHub-specific and npm-specific types of projects as application-specific and library-specific packages. 
Details of this approach are explained in Section \ref{subsec:rq2_analysis}.

%-------------------
\begin{table}[t]
  \centering
  \caption{Summary of Extracted Datasets.} \label{tab:filtered_packages}
  \begin{tabular}{ll|l|r}\hline
   \multicolumn{3}{l|}{Extraction Snapshot of Projects} & July 2008 - July 2016 \\ \hline
   \multicolumn{3}{l|}{Extracted \rmd files} & 153,857 packages \\ \cline{2-4}
%   Extracted Headlines & 79,898 headline variations \\ \hline
    & \multicolumn{2}{|l|}{\rmd files after preprocessing} &  43,911 packages \\ \cline{2-4}
    & \multicolumn{2}{|l|}{Content themes} & 30,939 content themes \\ \hline
%    & & npm-strong packages & 2,788 packages \\ \cline{3-4}
%    & & npm-strong contents & 2,254 contents \\ \cline{3-4}
%    & & GitHub-strong packages & 1,870 packages \\ \cline{3-4}
%    & & GitHub-strong contents & 1,567 contents \\ \hline
  \end{tabular}
\end{table}
%-------------------

%GitHubで開発されるプロジェクトのうち，JavascriptのエコシステムであるNPMパッケージリポジトリに登録されるプロジェクトを対象とする．NPMパッケージリポジトリは，2009年以降に急速にパッケージ登録数が増加し\cite{npm}，jQueryのようなライブラリから，expressのようなフレームワーク，gruntのようなコマンドラインツールまで，幅広い種別のプロジェクトのパッケージが管理されている．
%また各パッケージにはプロジェクトが発見されやすくするために，パッケージの種類を表すような``browser''や``cli''といったキーワードが割り振られており，プロジェクトの種別に応じた記述項目の違いを調査するために有用な分析対象である．割り振られたキーワードはプロジェクトがNPMパッケージリポジトリに登録する際に作成するpackage.jsonから確認できる．
%GitHubで開発されるプロジェクトのうち，JavascriptのエコシステムであるNPMパッケージリポジトリに登録されるプロジェクトは，2016年7月1日時点で153,838件のパッケージを確認した．本論文では，Witternら~\cite{npm}のデータセットの収集方法に基づいて取得した153,857件のリポジトリが公開するREADMEを分析対象とする．
%
%さらにそのうち，項目の抽出を容易にするために，Markdown記法で記述されているREADMEを対象とする．具体的には.md，あるいは.markdownという拡張子を持つREADMEを対象とする．拡張子を含め，大文字か小文字かの違いは考慮せず，どちらも同じファイル名を捉える．GitHubではMarkdown形式の他に.textileや.rdocといった拡張子のREADMEにも対応しているが，本論文では対象としない．
%Markdown記法で記述されているREADMEを持つプロジェクトは，143,933件（93.6\%）あった．
%
%単語の意味をなるべく一意とするため，英語で記述されたREADMEを対象とする．具体的には，READMEの見出しに記述された内容の文字コードをPythonのunicodedataライブラリを用いて判別し，漢字，ひらがな，カタカナ，キリル，ハングルの文字を持つREADMEを除外した．
%英語で記述されたREADMEを持つプロジェクトは141,685件あった．
%
%安定したプロジェクトを対象とするため，READMEを作成して3年以内であり，1年間更新していないものを対象とした．
%前者は開発期間が長すぎる，あるいは短すぎるプロジェクトを切り捨てる条件であり，後者は更新していないことをREADMEが完成し更新する必要がなくなったものと判断したことによる条件である．
%READMEを作成して3年以内であり，1年間更新していないREADMEを持つプロジェクトは43,991件あった．
%
%43,991件を対象に，各RQを調査する．

%Section3
%%%%%%%%%%%%%%%%%%%%%%%%%%%%%
\section{Empirical Study} \label{sec:rq}
%%%%%%%%%%%%%%%%%%%%%%%%%%%%%
In this section, we evaluate the two research questions proposed in Section \ref{subsec:rqs}. 
For each research question, we describe the approach and their results.

\subsection{\textbf{RQ1: \rqone}} \label{sec:rq1}

%----------------
\begin{table}[t]
  \centering
  \caption{Example of our cleaning (i.e., using stemming) of the headlines. In this data processing, we are able to map the variations of headlines to the install content theme} \label{tab:result_stemming_headline}
  \begin{tabular}{l|r}\hline
   Headline Variations &  Content Theme \\ \hline\hline
   How to Install? & \multirow{6}{*}{Install} \\ \cline{1-1}
   installing & \\ \cline{1-1}
   Install it &  \\ \cline{1-1}
   Installation & \\ \cline{1-1}
   INSTALL &  \\ \cline{1-1}
   1. Installing & \\ \hline
  \end{tabular}
\end{table}

\begin{table}[t]
  \centering
  \caption{Example of stopword exceptions in the mapping headline to content theme (Step 2)} \label{tab:rule_stopword_headline}
  \begin{tabular}{l|l|r}\hline
   Stopword token & Content Theme & Example of Headline \\ \hline\hline
   to do & todo & To do \\ \hline
   how & usage & How to \\ \hline
   who & \multirow{2}{*}{author} & Who are we? \\ \cline{1-1}\cline{3-3}
   from & & From \\ \hline
   more & \multirow{2}{*}{document} & More.... \\ \cline{1-1}\cline{3-3}
   other & & Others \\ \hline
   about & \multirow{4}{*}{overview} & ABOUT \\  \cline{1-1}\cline{3-3}
   what & & What the... ?\\ \cline{1-1}\cline{3-3}
   that & & What is that? \\ \cline{1-1}\cline{3-3}
   can & & What can I do? \\ \hline
  \end{tabular}
  \vspace{-2mm}
\end{table}
%----------------

\paragraph{\underline{\textbf{Approach}}} \label{subsec:rq1_analysis}
To answer RQ1, we perform an analysis of the \rmd file contents.
Our analysis consists of two steps:\\
\begin{itemize}
\item \textit{(Step 1) Extraction of \rmd content themes.} Our key assumption is that headlines in the \rmd file are indicative of important content themes. 
Since we find that \rmd files often use a header as a summary of their content themes, we perform the following: \\
\textit{(Step 1a) Extracting Headlines.} Targeting the levels 1 and 2 (i.e., $h1$ and $h2$) headlines, we extract 79,898 headlines using this technique.
Hence, using the markdown format, we can extract headlines using the syntax ($h1$: $\texttt{\#}$ and $\texttt{\#\#}$ and $h2$: $\texttt{===}$, $\texttt{---}$).\\
\textit{(Step 1b) Mapping Headline to Content.} Since a headline is a natural language, variations, and spelling inconsistencies cause noise in the dataset.
For example, developers use \content{How to Install?}, \content{installing}, \content{Installation} to summaries the \content{Install} content theme.
We use the stemming technique from the language processing package (i.e., \algorithm{nltk} package in Python) to normalize and clean noise in the data.
The \algorithm{nltk} package is well-known and provides a high accuracy of software engineering datasets~\cite{Omran_MSR2017}.\\
% %Stemming is performed as follows: 
% %(i) Remove headline or subheadline with single character in utf-8,
% %(ii) Split headline or subheadline into tokens and converts the letters to lower case,
% %(iii) Remove the tokens with only number and remove stop words, and
% %(iv) Make an item by connecting the rest tokens with ``\_''
%\end{enumerate}\\\
\textit{(Step 1c) Merging Similar Content Themes.} To further reduce the noise in the content theme dataset, we merge content themes that contain manually merged content themes with similar or related meanings. 
For example,  we conclude that content themes \content{Getting Started}, \content{To setup}, and \content{Download} should be merged into the \content{Install} content theme. In this study, the first author, second author, and third author firstly make clusters to merge content themes with each other. Next, if there are the content themes in the different clusters between the authors, we start a discussion to reach consensus on common content themes.
\item \textit{ (Step 2) Classification of \rmd content themes.} Based on the results of Step 1, we display the frequency count of each content theme and its coverage (i.e., the percentage of systems using each content theme). 
\end{itemize}

Using our approach, we extracted 30,939 content themes from 69,869 headlines from the \rmd files (i.e., Steps 1a and 1b).
Table~\ref{tab:result_stemming_headline} shows an example of Step 2.
Furthermore, Table~\ref{tab:rule_stopword_headline} shows some exceptions and adjustments that we encountered to the conventional natural language stemming approach.
For instance, using the default settings, the headline \content{who are we?} would be removed (i.e., Step 1b).

\pagebreak
By merging the more frequent content themes and filtering out the less common content themes (Step 1c and Step 2), we finally ended up with the 22 most frequent content themes used by more than 1\% of projects use from the 30,939 content themes(i.e., Step 1c). 
In a semi-automatic approach, we incrementally filter content themes not frequently appearing (i.e., Step 2).

\paragraph{\underline{\textbf{Result}}} \label{subsec:rq1_result}

%RQ1に答えるために，私たちは試行錯誤されたREADMEに頻繁に記述されている項目を，その件数と割合，項目に記述される内容，統合される前の見出し例とともに表\ref{tab:top_headline}に示す．
%As a result, we make the following observations:

%--------------------------
\begin{table*}[t]
  \centering
  \caption{A taxonomy of common content themes in \rmd files (note that PR=Packages/All packages). Note that we use the Top 6 content themes for analysis in $RQ_3$} \label{tab:top_headline}
  \scalebox{0.6}[0.6]{
  \begin{tabular}{c|p{2.5cm}|r|c|p{4.5cm}|p{5.5cm}}\hline
   \hfil \multirow{2}{*}{Rank} \hfil & \hfil Merged content \hfil & \hfil \multirow{2}{*}{Packages (PR)} \hfil & \hfil Guideline \hfil & \hfil \multirow{2}{*}{Description} \hfil & \hfil \multirow{2}{*}{Content Theme} \hfil \\
    & \hfil theme \hfil & \hfil  \hfil & \hfil from Table~\ref{tab:guideline} \hfil &  & \\ \hline\hline
1 & Usage & 26,758 (60.83\%) & \checkmark & Basic usage example of the project & Usage, Basic Usage, How to Use, Use Case, Methods, Screenshot, Examples, Quick Examples, Tips, Syntax, Sample, Hello World, Grunt tasks \\ \hline
2 & Install & 26,142 (59.43\%) & \checkmark & How to install the project & How to Install?, Installation,  Getting Started, Get started now!, To setup, Download, Initialization, Instructions, npm, node.js \\ \hline
3 & License & 15,932 (36.22\%) & \checkmark & Type of license applied to the project & LICENSE, MIT License, Unlicense, License, Copyright, Legal \\ \hline
4 & API & 10,675 (24.27\%) & & API list of the project & API, API References, API Documents, Command Line, CLI, Build, Events, Constructor, Action, To run, Objects, Interface, Commands, Function, Execute \\ \hline
5 & Option (Product) & 10,459 (23.78\%) & \checkmark & Option list of the project & Options, Format, Style, Parameter, Configure, Import, Custom, Compatibility, Browser, Config, Client, Server, Module, Promise, Util, Class, Variables, Router \\ \hline
6 & Release history & 5,874 (13.35\%) & \checkmark & Release history of the project & Release History, ChangeLog, Change logs, Version history, Versions, Release Notes \\ \hline
7 & Contribute & 4,938 (11.23\%) & \checkmark & How to contribute to the project & How to Contribute, Contribution Guides, Development, Donations, CONTACT, Hacking \\ \hline
8 & Test & 4,374 (9.94\%) & & Test commands of the project & Run test, testing, To test, How to Test \\ \hline
9 & Todo & 4,293 (9.76\%) & & TODO list of the project & TODO, Todos, To-Do List, To-do soon, Task, In the future ..., The future, Requirements, Coming soon! \\ \hline
10 & Overview & 4,219 (9.59\%) & \checkmark & Description of the project & Overview, Summary, Synopsis, Description, Introduction, Why?, What's this?, About The Name \\ \hline
11 & Status & 3,491 (7.94\%) & & Build status by continuous integration & Build Status, Current Status \\ \hline
12 & Document & 3,384 (7.69\%) & & Further Documentation for using the project & Documentation, Doc, Doe, Notes, Info, Information, TL;DR, Notice, Detail, See also \\ \hline
13 & Author & 2,607 (5.93\%) & \checkmark & Project author & Authors, About the Author, who am i?, Credits, Backers, Contributors, Other Contributors, Resources \\ \hline
14 & Support & 1,555 (3.53\%) & \checkmark & How to get support & Supports, FAQ?, Help, Troubleshooting, Questions \\ \hline
15 & Feature & 1,379 (3.13\%) & & Features list of the project & Key Features, Attributions \\ \hline
16 & Relate & 1,297 (2.95\%) & & Introduction of other projects related to the project  & Related, Related Projects, Link, Inspirations, Alternatives, Source, Other libraries \\ \hline
17 & Issue & 1,039 (2.36\%) & & Issues & Issues, Known issues, Problems, Warning, Caveats, Bugs \\ \hline
18 & Demo & 839 (1.91\%) & & Example output of the project & Demo, Codepen demo, Example Output, Result \\ \hline
19 & Purpose & 821 (1.87\%) & & Purpose of the project & Purpose, Goals, Solution,  Motivation, Background, Concepts, Key Ideas, Philosophy, Rationale \\ \hline
20 & Refer & 772 (1.75\%) & & References and Acknowledge & References, Thanks, Special Thanks, Acknowledgements \\ \hline
%21 & Limit & 772 (1.75\%) & & Limitation that system cover  & Limitations, Implementation, Validations, Browser Support \\ \hline
%22 & Error & 456 (1.04\%) & & Error case that project outputs & Error, Filters \\ \hline
%23 & Roadmap & 281 (0.64\%) & & TODO and Release history & ROADMAP, Road Map, Work Progress \\ \hline
%24 & Table\_Content & 275 (0.63\%) & & Table of README content & Table of Content, Content \\ \hline
%25 & Model & 225 (0.51\%) & & Model and Design of the project & Models, MVC Model, UI Model \\ \hline
%26 & Benchmark & 200 (0.45\%) & & Benchmark of the project & Benchmarks, Benchmark / Performance, Performance, Performance wins \\ \hline
  \end{tabular}
  }
\end{table*}
%--------------------------

\textit{Observation 1 --- Many software projects (36.22\%-60.83\%) contain \content{Usage}, \content{Install} and \content{License} content themes in their \rmd files.}

Table~\ref{tab:top_headline} shows \rmd content themes which we merged with the different headlines. 
These results also coincided with the recommended guideline content themes (i.e., Table~\ref{tab:guideline}), which we show using a check mark (\checkmark).

We found that the top three \rmd content themes are \content{Usage} (i.e., 60.83\%), \content{Install} (i.e., 59.43\%), and \content{License} (i.e., 36.22\%).
However, we find that developers often use different variations of the same word to explain each content theme.
For instance, developers may use \content{Example}, \content{Hello World}, and \content{Grunt tasks} content theme keywords as \content{Usage}.
Furthermore, while we knew that \content{License} is a common header, only 36.22\% of systems note it in their \rmd files. 
We suspected that developers may place license information in other meta-files (i.e., package.json) or as a separate document itself.

\textit{Observation 2 --- Results confirm that \rmd content themes are targeted at its end users.} 

\content{Usage} and \content{Install} represent how to use the system for users.
The other content themes for users, \content{API} (24.27\%) and the \content{Option} (23.78\%) content theme explain the different list of functions for package usage.
When there are many functions in a system, developers often note the feature list of the system with \content{Feature} content (3.13\%). 
They may note \content{Support} (3.53\%), \content{Demo} (1.91\%), \content{Limit} (1.75\%), and \content{Error} (1.04\%) together as troubleshooting. 

The lesser documented content themes are more related to contributors' information \content{Contribute}, \content{Test}, \content{Todo}, \content{Issue}, and \content{Roadmap}.
We suspected that such content themes may indicate that the packages are not fully mature and require more development to become stable for end users. 

%\vspace{2mm}
\textit{Observation 3 --- \content{Overview}, \content{Author} and \content{Support} are rarely noted in \rmd files.}

While \content{Overview}, \content{Author} and \content{Support} are typically included in \rmd as shown in Table~\ref{tab:guideline}, they account for less than 10\% of the systems. 
We found that 9.59\% for \content{Overview}, 5.93\% for \content{Author}, and 3.53\% for \content{Support} still exist in the \rmd files. 
Furthermore, \content{Overview} is a more generic content theme while developers are less likely to include contributor content themes such as \content{Author} and \content{Support}.

%%%%%%%%%%%%%%%%%%%%%%%%%%%%%%%%%%%%%%%%%%
%\subsection{README headline contents according to projects}
%%%%%%%%%%%%%%%%%%%%%%%%%%%%%%%%%%%%%%%%%%
\subsection{\textbf{RQ2: \rqtwo}} \label{sec:rq2}

\paragraph{\underline{\textbf{Approach}}} \label{subsec:rq2_analysis}

To answer RQ2, we explore the relationship between the type of software project and its \rmd content themes.
In detail, as a case study, we compare two types of npm projects--- \textit{library-specific projects} (i.e., projects used by other applications as third-party libraries) and \textit{application-specific projects} (i.e., projects used by end users as applications) \cite{Wittern_MSR2016}.
Our approach consists of two steps: 

%\begin{enumerate}
\noindent\textit{(Step 1) Classification of \rmd based on project type.}\\
As defined by Wittern et al.~\cite{Wittern_MSR2016}, we define two types of npm projects: (a) GitHub-specific (i.e., identified by the keywords: \keyword{gruntplugin}, \keyword{gulpplugin}, \keyword{express}, \keyword{react}, \keyword{authentication}) and (b) npm-specific (i.e., identified by the keywords:  \keyword{util}, \keyword{array}, \keyword{buffer}, \keyword{string}, \keyword{file}).

\noindent\textit{(Step 2): Identification of \rmd content theme patterns.}\\
To identify common usage patterns between the two types of projects (i.e., GitHub-specific and npm-specific).
Using our content theme results from RQ1, we then use the Association Rule Mining technique~\cite{zhang2002association}\cite{han2000mining} to identify common usage patterns.
%\end{enumerate}

Association Rule Mining is a method to extract a relationship between two or more items as an association rule from the combination of a large number of items. 
The association rule is represented by a pre-condition, which is the \rmd content (RC), and a pre-condition (Pt) as follows: Let $C$ be a set of \rmd content themes (i.e., ``Install'', ``License'', ...) and $Pt$ refer to a set of project types (i.e., GitHub-specific or npm-specific).
\begin{equation} 
\mathbb{R} = RC \Rightarrow Pt
\end{equation}
To evaluate the extracted rules $\mathbb{R}$ , we use the three metrics: support, confidence, and lift. 
We define the support as the proportion of rules where both pre-condition (RC) and post-condition (Pt) exist in all rules. 
\begin{equation} 
%\scalebox{0.8}{$\displaystyle
support(RC)=\frac{|\sigma(RC\cap Pt)}{|Pt|}
%$}
\end{equation}
The confidence metric is the proportion of rules which both the pre-condition (RC) and post-condition (Pt) exist in rules with the pre-condition (RC).
\begin{equation} 
%\scalebox{0.8}{$\displaystyle
conf(\mathbb{R}) 
=\frac{support(RC \cup Pt)}{support(RC)}
%$}
\end{equation}

Finally, Lift measures the magnification of the data in which the pre-condition (RC) and post-condition (Pt) exist in rules with the post-condition (Pt).

\begin{equation} 
%\scalebox{0.8}{$\displaystyle
lift(\mathbb{R})=\frac{conf(\mathbb{R})}{support(Pt)}
%$}
\end{equation}

Our study implements association rule mining using the \algorithm{Orange}~\cite{JMLR:v14:demsar13a} library in Python. 
The library uses an apriori algorithm, which is used to filter out the minor rules (i.e., using minimum support value (0.03), minimum confidence value (0.03), and minimum lift value (1)).

Table \ref{tab:filtered_github_npm_packages} shows the identified \rmd files (Step 1). 
In fact, we extracted 2,788 GitHub-specific packages and 1,870 npm-specific packages from our dataset. 
During the analysis, we discarded the keyword \keyword{gruntplugin} as gruntplugin related projects use different \rmd format. We also ignore systems which include keywords from both project types.

%----------------------
\begin{table}[t!]
  \centering
  \caption{GitHub-specific packages and NPM strong packages} 
  \label{tab:filtered_github_npm_packages}
  \scalebox{1.0}[1.0]{
  \begin{tabular}{l|r}
  \multicolumn{2}{l}{GitHub-specific packages} \\\hline
   Filtered Projects 	& 	2,788 packages \\ \hline
   Content Theme 	& 	2,254 kinds of headline variations \\ \hline
   \multicolumn{2}{l}{}\\ 
   \multicolumn{2}{l}{npm-specific packages} \\\hline
   Filtered Packages 	& 	1,870 packages \\ \hline
   Content Theme	& 	1,567 kinds of headline variations \\ \hline
  \end{tabular}
	}
\end{table}
%----------------------

We report our \rmd content theme patterns in two forms.
Table~\ref{tab:apriori_rule} shows the extracted 36 rules which the target README files frequently use sorted by lift value. 
The second representation is a graph-based visualization of the generated rules. 

%---------------------------
\begin{table*}[t!]
\centering
\caption{Top 36 Association Rules for GitHub-specific vs. \colorbox{gray!25}{npm-specific packages}.} \label{tab:apriori_rule}
	\scalebox{0.9}[0.9]{
  	\begin{tabular}{rlcl|ccc}
\hline
Id &     Content Themes	&	$\Rightarrow$	&	 Project type  &      support  &  confidence &lift \\ \hline
\rowcolor{gray!25} 1 &  \{Usage,Install,License,Test\}	&	$\Rightarrow$	&	\{npm-specific\}	&	0.04	&	0.60	&	1.49	\\
\rowcolor{gray!25} 2 &  \{Install,License,Status\}	&	$\Rightarrow$	&	\{npm-specific\}	&	0.03	&	0.59	&	1.48	\\
\rowcolor{gray!25} 3 &  \{Install,License,Test\}	&	$\Rightarrow$	&	\{npm-specific\}	&	0.05	&	0.59	&	1.47	\\
\rowcolor{gray!25} 4 &  \{License,Status\}	&	$\Rightarrow$	&	\{npm-specific\}	&	0.04	&	0.58	&	1.44	\\
\rowcolor{gray!25} 5 &  \{Usage,License,Test\}	&	$\Rightarrow$	&	\{npm-specific\}	&	0.04	&	0.58	&	1.43	\\
\rowcolor{gray!25} 6 &    \{License,Test\}	&	$\Rightarrow$	&	\{npm-specific\}	&	0.05	&	0.57	&	1.42	\\
\rowcolor{gray!25} 7 &  \{Install,Status\}	&	$\Rightarrow$	&	\{npm-specific\}	&	0.04	&	0.56	&	1.41	\\
\rowcolor{gray!25} 8 &  \{Usage,Install,Test\}	&	$\Rightarrow$	&	\{npm-specific\}	&	0.05	&	0.56	&	1.39	\\
\rowcolor{gray!25} 9 &  \{Usage,Status\}	&	$\Rightarrow$	&	\{npm-specific\}	&	0.04	&	0.54	&	1.35	\\
\rowcolor{gray!25} 10 & \{Install,Test\}	&	$\Rightarrow$	&	\{npm-specific\}	&	0.06	&	0.54	&	1.35	\\
\rowcolor{gray!25} 11 & \{Install,License,API\}	&	$\Rightarrow$	&	\{npm-specific\}	&	0.06	&	0.54	&	1.34	\\
\rowcolor{gray!25} 12 & \{Usage,Test\}	&	$\Rightarrow$	&	\{npm-specific\}	&	0.06	&	0.54	&	1.34	\\
\rowcolor{gray!25} 13 & \{Status\}	&	$\Rightarrow$	&	\{npm-specific\}	&	0.05	&	0.53	&	1.33	\\
\rowcolor{gray!25} 14 & \{Test\}	&	$\Rightarrow$	&	\{npm-specific\}	&	0.07	&	0.53	&	1.31	\\
15 & \{Usage,Option\}	&	$\Rightarrow$	&	\{GitHub-specific\}	&	0.14	&	0.76	&	1.27	\\
16 & \{License,Option\}	&	$\Rightarrow$	&	\{GitHub-specific\}	&	0.08	&	0.75	&	1.25	\\
\rowcolor{gray!25} 17 & \{Install,API\}	&	$\Rightarrow$	&	\{npm-specific\}	&	0.09	&	0.50	&	1.25	\\
18 & \{Install,Option\}	&	$\Rightarrow$	&	\{GitHub-specific\}	&	0.12	&	0.74	&	1.24	\\
19 & \{Option\}	&	$\Rightarrow$	&	\{GitHub-specific\}	&	0.19	&	0.73	&	1.21	\\
\rowcolor{gray!25} 20 & \{License,API\}	&	$\Rightarrow$	&	\{npm-specific\}	&	0.07	&	0.47	&	1.18	\\
\rowcolor{gray!25} 21 & \{Install,License\}	&	$\Rightarrow$	&	\{npm-specific\}	&	0.16	&	0.47	&	1.17	\\
\rowcolor{gray!25} 22 & \{Usage,Install,Releas History\}	&	$\Rightarrow$	&	\{npm-specific\}	&	0.03	&	0.46	&	1.16	\\
\rowcolor{gray!25} 23 & \{Install,Author\}	&	$\Rightarrow$	&	\{npm-specific\}	&	0.03	&	0.46	&	1.16	\\
\rowcolor{gray!25} 24 & \{Install,Releas History\}	&	$\Rightarrow$	&	\{npm-specific\}	&	0.04	&	0.46	&	1.14	\\
25 & \{Todo\}	&	$\Rightarrow$	&	\{GitHub-specific\}	&	0.04	&	0.68	&	1.13	\\
\rowcolor{gray!25} 26 & \{API\}	&	$\Rightarrow$	&	\{npm-specific\}	&	0.11	&	0.45	&	1.13	\\
27 & \{Usage,Overview\}	&	$\Rightarrow$	&	\{GitHub-specific\}	&	0.04	&	0.67	&	1.11	\\
\rowcolor{gray!25} 28 & \{License\}	&	$\Rightarrow$	&	\{npm-specific\}	&	0.20	&	0.44	&	1.10	\\
29 & \{Install,Overview\}	&	$\Rightarrow$	&	\{GitHub-specific\}	&	0.04	&	0.65	&	1.09	\\
30 & \{Overview\}	&	$\Rightarrow$	&	\{GitHub-specific\}	&	0.05	&	0.65	&	1.08	\\
\rowcolor{gray!25} 31 & \{Author\}	&	$\Rightarrow$	&	\{npm-specific\}	&	0.04	&	0.43	&	1.07	\\
\rowcolor{gray!25} 32 & \{Install\}	&	$\Rightarrow$	&	\{npm-specific\}	&	0.25	&	0.42	&	1.05	\\
\rowcolor{gray!25} 33 & \{Releas History\}	&	$\Rightarrow$	&	\{npm-specific\}	&	0.05	&	0.42	&	1.04	\\
\rowcolor{gray!25} 34 & \{Contribute\}	&	$\Rightarrow$	&	\{npm-specific\}	&	0.04	&	0.41	&	1.03	\\
35 & \{Usage\}	&	$\Rightarrow$	&	\{GitHub-specific\}	&	0.41	&	0.61	&	1.02	\\
36 & \{Document\}	&	$\Rightarrow$	&	\{GitHub-specific\}	&	0.05	&	0.61	&	1.01	\\
\hline
\end{tabular}
}
\end{table*}

\begin{figure*}[t]
\captionsetup{width=6.5cm}
  \begin{center}
  \includegraphics[width=6.5cm,clip]{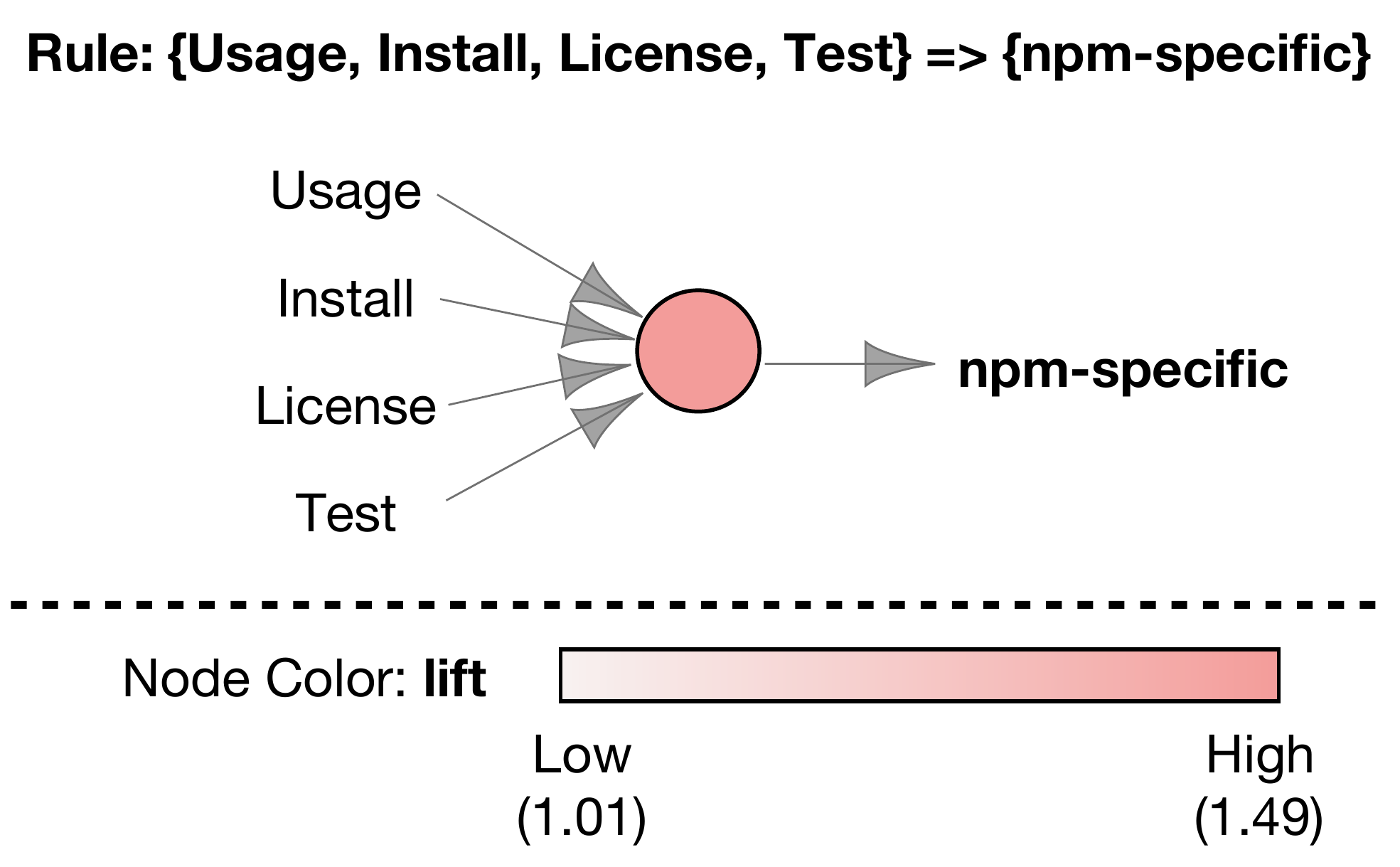}
  \end{center}
    \caption{An example of the content theme pattern rule represented as a directed graph, with nodes and edges. Note that the color indicates the lift metric.}
    \label{fig:exampleAR}
  \begin{center}
  \includegraphics[width=8cm,clip]{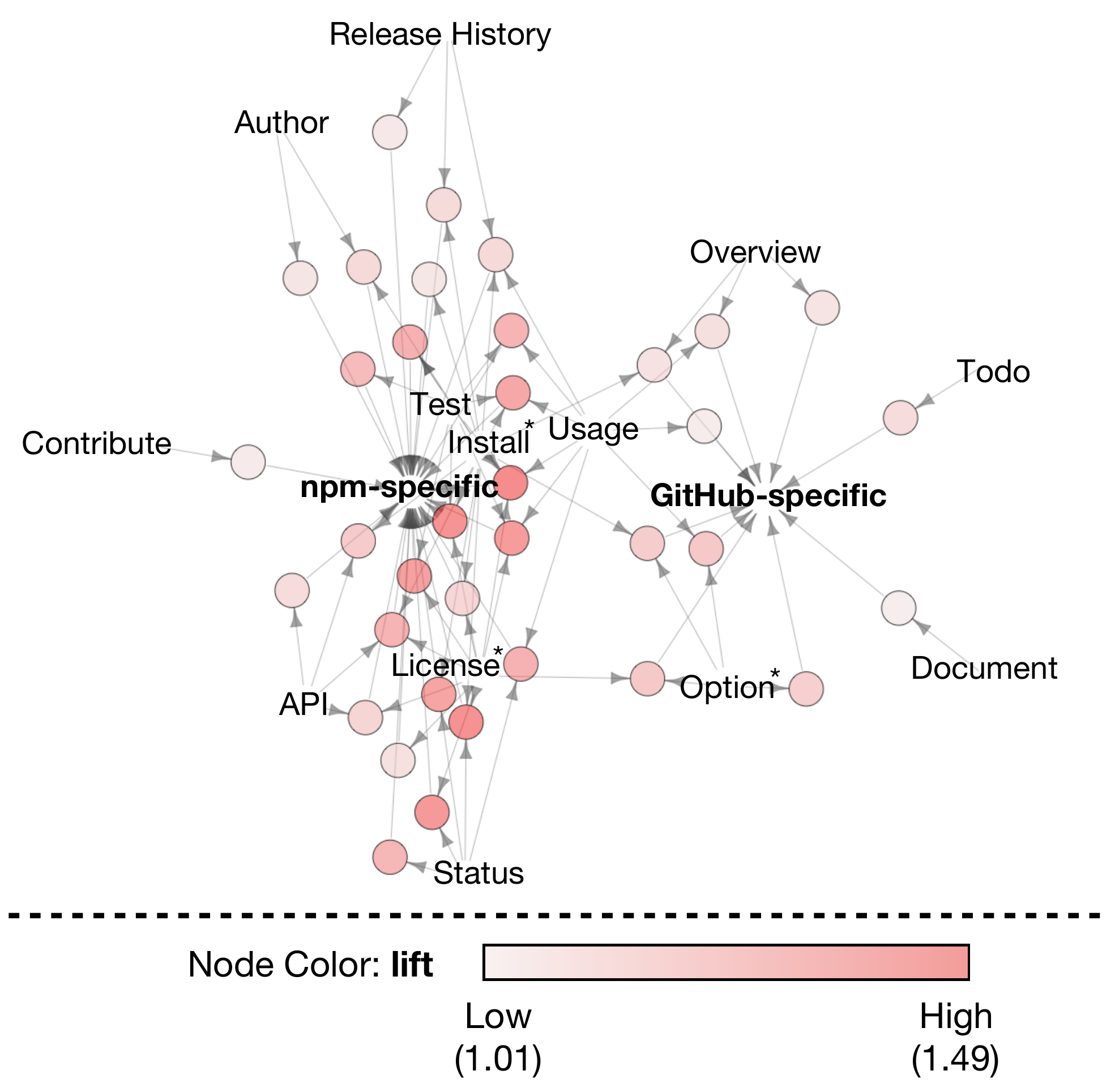}
  \end{center}
\caption{36 \rmd content theme rules generated as a directed graph.}
\label{fig:apriori_rule_graph}
\end{figure*}
%---------------------------

Figure \ref{fig:exampleAR} shows how our generated rule is translated into a graph representation. 
In this example, each of the incoming edges (i.e., \content{Usage}, \content{Install}, \content{License}, \content{Test}) represents each of the precondition content themes, while the outgoing edge is the postcondition (i.e., npm-specific)
The color of the node represents the lift metric, and as shown in the example, the color shows 1.49 lift.
Our assumption is that the colors will show whether or not the content theme is related to either ``Github-specific'' or ``npm-specific''.

Figure~\ref{fig:apriori_rule_graph} shows the generated graph (i.e., displayed using the \algorithm{Fruchterman Reingold} algorithm) for the 36 rules generated in Table~\ref{tab:apriori_rule}.
The color of the node follows the lift score from a light color (low) to a strong color (high). 

\vspace{2mm}
\paragraph{\underline{\textbf{Result}}} \label{subsec:rq2_result}

\textit{Observation 1 --- \content{Install} and \content{License} content themes are likely to be important for a npm-specific package.}

While $RQ_1$ shows that a \rmd file typically includes \content{Usage}, \content{Install} and \content{License}, Figure~\ref{fig:apriori_rule_graph} provides evidence that the \content{License} content theme is closely related to the npm-specific type of packages. 
Correspondingly, Table~\ref{tab:apriori_rule} shows that many npm packages with higher lift and higher confidence scores have some of the same rules as \content{License}. 
From this result, we believe that it is important for core utility packages to share the license because it is frequently reused by the other systems. 
Additionally, the graph shows that the \content{Status} and \content{Test} content themes are more closely associated closely to npm-specific types of libraries.
Furthermore, we suspect that the developers of these packages may be expected to be adapted correctly to the `test' and the `build' states.

\vspace{2mm}
\textit{Observation 2 --- The \content{Option} content theme is likely to be an important content theme for GitHub-specific packages.}
Figure~\ref{fig:apriori_rule_graph} shows that \content{Option} is closely related to the GitHub-specific type of packages. 
Table~\ref{tab:apriori_rule} shows that many of the GitHub-specific packages have higher lift and higher confidence scores with rules associated with the \content{Option} content theme. 
Furthermore, we suspect that the \content{Option} content themes are more important for the end-user package, as end user packages are more likely to have more lists of product options than npm-specific type (i.e., utility) packages.

%%% Section 4
%%%%%%%%%%%%%%%%%%%%%%
%%%%%%%%%%%%%%%%%%%%%%
\section{Summary of Results} \label{sec:discussion}
%%%%%%%%%%%%%%%%%%%%%%

In order to aid developers faced with documentation issues, we conducted an empirical study to understand the written content themes of the \rmd file.
The co-founder of GitHub Tom Preston-Werner, even discussed the importance of the \rmd file, coining Readme Driven Development (RDD)\footnote{Readme Driven Development: \url{http://tom.preston-werner.com/2010/08/23/readme-driven-development.html}} as an important subset of Document Driven Development.
We learned some valuable lessons along the way:

\begin{itemize}
\item\textit{Lesson 1:} \textit{Although a \rmd file contains numerous variations, we built a taxonomy of 22 \rmd content themes} -
Surprisingly, from over 30,000 content theme variations, we were able to build a taxonomy of 22 headline content themes, which are used by more than 1\% of packagess. 
We conjecture that the content themes may reveal insights such as project practices or may be an indicator of changes in the project.
\vspace{1mm}
\item\textit{Lesson 2: Content themes \content{Usage}, \content{Install}, and \content{License}  are common \rmd content themes} - \content{Usage}, \content{Install}, and \content{License} are typically included in the \texttt{README}. 
Furthermore, less apparent \rmd content themes include \content{API}, \content{Test}, and \content{Todo}, used in 10\%-24\% of packages.
Such information may be important, especially for novice developers.
\vspace{1mm}
\item\textit{Lesson 3:} \textit{Especially for npm packages, the study shows that \content{Install} and \content{License} are likely content themes for library-specific packages, while the \content{Option} content theme is more common for application-specific packages.}\\
We found some specific README content themes according to the type of projects.  We found \content{Install} (40\% packages) and \content{License} (20\% packages) are common for npm libraries, while nodeJS application packages included the option content themes (10\% packages).
Such information may be important for developers, especially for novice developers.
\end{itemize}

%%%%%%%%%%%%%%%%%%%%%%%
%\section{Discussion} \label{sec:discussion}
%%%%%%%%%%%%%%%%%%%%%%%
%In this section, we discuss the main findings of our study.
%%教訓．lesson
%
%%今回の結果の使い道．
%ガイドラインの具体化に貢献する：これまでREADMEはプロジェクトの開発者らによって経験的に作成されており，またガイドラインも実証的に有効とされたものはなかった．加えて，READMEの記述項目を多くのプロジェクトを横断して，実証的に調査した論文は存在しない．READMEはGitHubにおいて利用者や貢献者が初めに確認できるドキュメントであり，ほぼ全てのプロジェクトで利用されている重要なドキュメントであり，このREADMEにおける多くのプロジェクトが必要だと考える項目を明らかにできたことの貢献は大きいと考える．
%具体的にはガイドラインの通り，Usage，Install，Licenseは多くのプロジェクトが記述すべきだと考えていた．一方でContributeやSupportは多くのプロジェクトが記述していたわけではない．貢献を必要としないプロジェクトも存在し，ガイドラインが全てのプロジェクトを考慮していないことが確認できる．
%我々はプロジェクトの特徴に合わせて記述する項目が異なることを明らかにするとともに，異なる項目も明らかにした．例えば，コアユーティリティパッケージはTestやStatusを重視し，エンドユーザーアプリケーションはOptionを重視する．この事実は，プロジェクトの特徴に合わせたガイドラインの具体化を可能にする．
%またガイドラインでは曖昧だが，記述の方法は複数あることを明らかにした．例えばUsageだけでなく，より詳細な機能一覧（API，Option）や出力のデモ（Demo）などを記述するプロジェクトもあった．この解釈の違いを明らかにできたことは，記述すべき内容だけでなく，その具体的な方法も示唆可能なガイドラインの作成を支援すると考える．具体的なガイドラインにより，表記揺れも統一され，読みやすくなることが期待される．
%
%READMEに基づくプロジェクトの特徴判定：我々はプロジェクトの特徴ごとに記述する項目が異なることを明らかにした．加えてgruntpluginのように独自の形式で記述しているREADMEを発見した．例えばgruntpluginは'Task'という用語をUsageの記述に利用する．READMEにプロジェクトの特徴が現れることは，プロジェクトの特徴抽出にREADMEの項目を利用可能であることを意味すると考える，
%%クローンプロジェクトの判定の研究などで，プロジェクトの特徴抽出にREADMEの項目を利用可能である．こんな研究に役立つ．
%
%記述項目の推薦：我々はプロジェクトの特徴ごとに，後から追加されることになる項目を明らかにした．このことは記述する項目の推薦につながると考える．例えば，エンドユーザーパッケージであれば，Usageを後から追加することになることが多いので，READMEの作成直後に記述しておくべきといった推薦ができる．

%分析手法：
%項目を統合できる．この項目は似ている．
%今後の項目分析に有用である．

%ユーザの期待と開発者の期待：
%書けとされているのに書いていない．
%書けとされていないのに書いている．
%意識の差が明確になった．

%%% Section 5
%%%%%%%%%%%%%%%%%%%%%%%%%%%%%
\section{Threats to Validity} \label{sec:threats_to_validity}
%%%%%%%%%%%%%%%%%%%%%%%%%%%%%

\textit{External validity} - refers to the generalization concerns of the study to other software systems including some package ecosystem such as Java library and Ruby RubyGems. 
This study found some specific results for the npm package ecosystem. 
For example, while \content{Install} is reported as one a major content theme from our findings and is used by 59.43\% of systems, we carefully restrict these findings to the npm package ecosystem because other systems may depict different patterns and tendencies for a \rmd file.
This might be an interesting future avenues for research.

\textit{Internal validity} - refers to the concerns that are internal to this study. 
In this study, we found two main internal threats that could affect our results. 
First is the preprocessing of the dataset. 
In $RQ_1$, we classified 30,939 content themes into the 22 most frequent content themes by merging the more frequent contents and filtering out the less common content themes. 
The manual merging of content themes in $RQ_1$ was conducted through a reached consensus among authors. 
However, we followed a strict iterative process and are confident of the results.
The second threat is related to the content themes of the \rmd files (i.e., $RQ_2$). 
As shown in our results, not every \rmd file will include key content themes.
For example, some projects have separate meta-files for licenses; thus, the content theme for licenses may not exist in the \rmd file.
For future work, it will be interesting to investigate all meta-files to understand how developers maintain and keep all files.

\textit{Construct validity} - refers to the concerns of the result.
We found one threat that related to the extraction of content themes from the \rmd file. 
This study used the Markdown Format to extract the headline levels 1 and 2 (i.e., \textit{h1} and \textit{h2}). 
There may, however, be cases where the project is using level 3 (i.e., \textit{h3}) to write major content themes. 
Nonetheless, we are confident of the results and of our extraction approach.

%%% Section 6
%%%%%%%%%%%%%%%%%%%%%%%%%%%%%%
\section{Conclusions and Future work} \label{sec:conclusion}
%%%%%%%%%%%%%%%%%%%%%%%%%%%%%%
In this paper, we investigated content themes of the \rmd file.
Although we found that the \rmd file contains numerous ambiguous naming variations, we were able to summarize and build a taxonomy of 22 \rmd content themes used by more than 1\% of packages. 
The results show that \rmd files contain common content themes such as \content{Usage}, \content{Install}, and \content{License}, as outlined in known guidelines. 
Furthermore, we found that \content{Install} and \content{License} are likely content themes for library-specific packages, while the \content{Option} content theme is more common for application-specific packages.
Finally, we showed that packages rarely remove \rmd content themes.
    
As future work, we would like to extend our project types and techniques to provide more comprehensive guidelines for writing a good \rmd file.
We also believe that further understanding of \rmd will assist both developers and their end users in keeping up with ongoing changes in a project.

\section*{Acknowledgments}
%\begin{acknowledgements}
%\end{acknowledgements}

\bibliographystyle{spmpsci}      % mathematics and physical sciences
\bibliography{bibfile}   % name your BibTeX data base

\end{document}